\pdfoutput=1

\documentclass[aps,twocolumn,amsmath,amssymb,preprintnumbers,floatfix,prb,superscriptaddress,longbibliography]{revtex4-2}

\usepackage[utf8]{inputenc}
\usepackage{newtxtext}
\usepackage[upint]{newtxmath}
\usepackage{microtype}
\usepackage{textcomp}
\usepackage{eucal}
\usepackage{bm}
\usepackage{siunitx}
\usepackage{comment}

\usepackage{enumerate}
\usepackage{amsfonts}
\usepackage{amsmath}
\usepackage{amssymb}
\usepackage{color}
\usepackage{soul}

\usepackage{graphicx}

\usepackage[colorlinks,allcolors=blue]{hyperref}
\usepackage[capitalize]{cleveref}

\definecolor{DarkRed}{rgb}{0.65,0,0}%
\definecolor{Green}{rgb}{0,0.3,0.3}
\definecolor{Purple}{rgb}{0.3,0,0.65}
\definecolor{Red}{rgb}{1,0,0}
\definecolor{Blue}{rgb}{0,0,0.85}
\definecolor{Magenta}{rgb}{1,0,1}

\newcommand{\hans}[1]{\textcolor{Magenta}{{#1}}}

\newcommand{\Imag}{{\Im\mathrm{m}}}   
\newcommand{\Real}{{\mathrm{Re}}}   
\newcommand{\ve}[1]{\boldsymbol{#1}}
\DeclareMathOperator{\diag}{diag} 

\newcommand{\vecA}{\ve{A}} 

\newcommand{\e}[1]{\mathrm{e}^{#1}}
\newcommand{\ubar}[1]{\underline{#1}}

\newcommand{\vecp}{\ve{p}}

\newcommand{\vecB}{\ve{B}}

\def\i{\mathrm{i}}

\newcommand{\be}{\begin{equation}}
\newcommand{\ee}{\end{equation}}

\renewcommand{\vec}[1]{\boldsymbol{#1}} 
\newcommand{\Tr}[1]{\text{Tr}\! \left\{ #1\right\}}
\newcommand{\sign}[1]{\text{sgn}\! \left(#1\right)}

\usepackage{pgfplots}
\pgfplotsset{compat=1.5, width=6.cm}
\usepgfplotslibrary{external}
\usepackage{subcaption}

\newcommand{\prlsection}[1]{\textit{#1}.\kern0.05em---\kern0.05em\ignorespaces}

\begin{document}
\title{Effective quasiclassical models for odd-frequency superconductivity: \\
energy-symmetry, preserved spectral weight, and Meissner response}
\author{Hans Gløckner Giil}
\affiliation{Center for Quantum Spintronics, Department of Physics, Norwegian \\ University of Science and Technology, NO-7491 Trondheim, Norway}
\author{Jacob Linder}
\email[Corresponding author: ]{jacob.linder@ntnu.no}
\affiliation{Center for Quantum Spintronics, Department of Physics, Norwegian \\ University of Science and Technology, NO-7491 Trondheim, Norway}

\begin{abstract}
The odd-frequency superconducting state appears generally in hybrid structures consisting of conventional superconductors and other materials, and features electrons that form temporally non-local Cooper pairs. The quasiclassical theory of superconductivity has been extensively used to model such systems, finding in many cases excellent agreement with experimental measurements. Therefore, it is of interest to study effective models of odd-frequency superconductivity to predict new phenomena associated with this form of unconventional pairing. 
We establish necessary criteria that the quasiclassical Green functions in odd-frequency superconducting systems in the dirty limit must satisfy in order to be physically reasonable, including conservation of spectral weight. We show that it is possible to write down effective models which satisfy all the abovementioned criteria, but which still predict different behavior when it comes to the density of states and the magnetic response of the superconductor. For instance, an effective model for the odd-frequency anomalous Green function that gives a conserved spectral weight can yield either a peaked or gapped density of states at the Fermi energy, and exhibit conventional, zero, or unconventional Meissner response. This finding demonstrates the importance of carefully considering the properties of effective models describing odd-frequency superconductivity in order to obtain physically reasonable results.
\end{abstract}
\maketitle

\section{Introduction}

In the theory of superconductivity, symmetry is a key component which dictates several properties of the elementary building block of superconductors: the Cooper pair. The function mathematically describing how the two electrons making up the Cooper pair correlate to each other depends, even in the simplest case of a single-band superconductor, on the position, spin, and time coordinate of the electrons. In BCS theory \cite{BCS}, the time coordinate is usually ignored after introducing a cutoff for the energy range that the electrons attract each other within. Nevertheless, the symmetry property of a paired Cooper pair state allows for the intriguing possibility that the two electrons are not correlated at equal times and that their correlation is instead established as the time separation grows. This is the case if the correlation function is odd in the relative time coordinate of the electrons.

Berezinskii \cite{berezinskii} predicted that a two-electron pairing correlation, with temporal coordinates $t_1$ and $t_2$, could be odd in the relative time coordinate $t_1-t_2$. An equivalent way of expressing this is to say that the time-ordered correlation function is odd in frequency, where the frequency is the Fourier-transform of the relative time coordinate.
Initially considered as an intrinsic pairing instability \cite{kirkpatrick_prl_91, belitz_prb_92, balatsky_prb_92, coleman_prl_93}, the field of odd-frequency superconductivity experienced a breakthrough when it was shown \cite{bergeret_prl_01} that such unconventional pairing could be generated by placing a conventional BCS superconductor in contact with a ferromagnet \cite{bergeret_rmp_05,buzdin_rmp_05}. Although no intrinsic bulk odd-frequency superconductors are known to date, it is by now well-established both theoretically and experimentally that such pairing exists in hybrid structures of essentially any sort featuring at least one conventional superconductor \cite{eschrig_jltp_07, tanaka_prl_07b, linder_rmp_19}.

Odd-frequency superconductivity has been shown to give rise to superconducting properties that differ fundamentally from those of conventional $s$-wave superconductors such as Al and Nb. This includes phenomena such as spin supercurrents \cite{keizer_nature_06,khaire_prl_10,robinson_science_10, eschrig_rpp_15}, paramagnetic Meissner effects \cite{dibernardo_prx_15, fominov_prb_15}, gapless superconductivity \cite{kontos_prl_01, bergeret_prb_02}, and the appearance of Majorana bound states \cite{asano_prb_13, huang_prb_15}. Because of this, it is of interest to consider effective models describing odd-frequency superconductivity in order to predict new emergent phenomena due to this interesting type of long-range order.

Previous works \cite{fominov_jetp_07, tanaka_prl_07, sukhachov_prb_19, johnsen_prb_21} have considered effective models for the anomalous Green function of an odd-frequency superconductor. Most of these utilized a methodology based on the quasiclassical theory of superconductivity \cite{eilenberger_zphys_68, usadel_prl_70, belzig_micro_99}, which is known to compare well with experimental measurements on mesoscopic metallic and superconducting systems. These models have satisfied the appropriate symmetries with respect to inversion of energy \cite{asano_prb_14}, fulfilling also the normalization condition $\check{g}^2=1$ of the quasiclassical Green function $\check{g}$ (some authors use a different convention with  $\check{g}^2 = -\pi^2$). However, a Green function satisfying all of these criteria can still display physically problematic behavior. 
This happens in at least two possible ways. One problem which previous works have suffered from is that the spectral weight of the system is not conserved \cite{tanaka_prl_07} when transitioning from the normal state to the superconducting state, i.e. a sum rule for the spectral weight is violated, as pointed out in Ref. \cite{sukhachov_prb_19}. Typically, spectral weight is experimentally found to be redistributed energetically when entering the superconducting state, but the number of states should remain the same. Moreover, as we will demonstrate below, an anomalous Green function which satisfies both the correct (i) symmetry relation under energy-inversion for odd-frequency superconductivity, (ii) a conserved spectral weight in the superconducting state, and (iii) a normalized quasiclassical Green function, can \textit{still} exhibit an arbitrary Meissner response: diamagnetic, paramagnetic, or even absent. In addition, it can give rise to either a gapped or peaked density of states at the Fermi level.

Because of these issues, we believe it is useful to clarify the precise form of the Green function which describes an odd-frequency superconductor which conserves the spectral weight of the normal-state, and to also predict which Meissner response and density of states it produces. Below, we outline four necessary, but not sufficient, criteria which a physically correct Green function describing an odd-frequency superconductor should satisfy. Then, we consider different models for such Green functions and show how they produce qualitatively different Meissner responses. 

\section{Theory}

\subsection{Preliminary: properties of odd-frequency Green function in S/F bilayers}

It is instructive to start by studying a superconductor/ferromagnet (S/F) system, in which odd-frequency arises naturally through the proximity effect.
We assume that the superconductor retains its BCS-solution, and consider the effect on the ferromagnet. 
The Usadel equation \cite{usadel_prl_70} holds in the diffusive limit and can, in the absence of electromagnetic fields and spin-flip and spin-orbit scattering, be written as 
\begin{equation}
    \label{eq:weak_usadel}
    D \partial_x \bigl(  \hat g \partial_x \hat g \bigr)  
    = -\i [E \hat \tau_3 + \hat M, \hat g],
\end{equation}
where $\hat M = \vec h \cdot \diag(\underline{\vec \sigma},\underline{ \vec \sigma}^*)$ is the ferromagnetic term with $\vec h$ being the spin-splitting energy, $\hat g$ is the retarded impurity averaged quasiclassical Green function, $\hat \tau_3 = \diag(1,1,-1,-1)$, $D$ is the diffusion constant of the system, $E$ is the energy relative to the Fermi energy, and $\underline{\vec \sigma}$ is a 3-vector containing the Pauli matrices. 
In this text, $\underline{A}$ denotes a $2 \times 2$ matrix in spin space, while $\hat A$ is a $4 \times 4$ matrix in spin-particle-hole space, also called spin-Nambu space \cite{nambu_pr_60}.
Finally, $\check A$ denotes an $8\times8$ matrix in Keldysh-spin-Nambu space.
The quasiclassical Green function has the form
\begin{align}
\check{g} = \begin{pmatrix}
\hat{g}^R & \hat{g}^K \\
\hat{0} & \hat{g}^A \\
\end{pmatrix},
\end{align}
where $\hat g^R$, $\hat g^A$, and $\hat g^K$ are the retarded, advanced, and Keldysh components. In the following we will consider the retarded Green function, dropping the suprescript.
The general structure of the retarded Green function contains the normal Green function $\underline g$ and the anomalous Green function $\underline{f}$, 
\begin{equation}
    \hat g = 
    \begin{pmatrix}
    \underline g & \underline f
    \\
    -\underline{\tilde f} & -\underline{\tilde g}
    \end{pmatrix}.
\end{equation}
Above, we have used the “tilde conjugation” operation, which for a function $A(E)$ is defined as
\begin{equation}\label{eq:oddfreqE}
    \tilde A(E) = [A(-E)]^*,
\end{equation}
and the $2\times 2$ matrix $\underline g$ has components \begin{equation}
    \underline g = 
    \begin{pmatrix}
    g_{\uparrow \uparrow} & g_{\uparrow \downarrow}
    \\
    g_{\downarrow \uparrow} & g_{\downarrow  \downarrow}
    \end{pmatrix},
\end{equation}
and similarly for the other matrices in spin space.

The superconducting order parameter $\Delta$ vanishes in the ferromagnet under the assumption of zero intrinsic attractive interaction without contact with the superconducting layer, and the anomalous Green function arises solely due to the proximity effect that exists due to tunneling between the materials.
In the weak proximity effect regime, where $\hat g$ takes a form close to the normal metal solution, we linearize the Usadel equation by writing
\begin{equation}
    \hat g = \hat \tau_3 + \delta \hat{f},
\end{equation}
and neglect terms that are second order in $\delta\hat{f}$. 

From the general symmetries of the quasiclassical Green functions, it can be shown that the retarded and advanced components are related as \cite{linder_rmp_19}
\begin{align}
    f_{\sigma \sigma'}^A(E) &= -f_{\sigma \sigma'}^R(-E), \text{ for triplet odd-frequency} 
    \\
    f_{\sigma \sigma'}^A(E) &=  + f_{\sigma \sigma'}^R(-E), \text{ for singlet even-frequency},
\end{align} 
where the superscripts indicate the retarded and advanced components of the Green function. In the considered regime of diffusive transport, the even-frequency correlations are spin-singlet whereas the odd-frequency correlations are spin-triplet.
Although these relations can be useful, they do nothing towards our goal of finding criteria for the retarded Green function itself. 
We will in the following derive symmetries for the retarded Green function directly.

Introducing a singlet-triplet decomposition according to $\delta f = (f_s +  \vec d \cdot \vec \sigma) \i \sigma_2$, where $f_s$ is the singlet component and $\vec d$ are the three triplet components,
the linearized Usadel equation can be written as a set of four coupled differential equations, 
\begin{align}\label{eq:linear}
    \frac{\i D}{2} \partial_x^2 f_s &= E f_s + \vec h \cdot \vec d
    \\
    \frac{\i D}{2} \partial_x^2 \vec d &= E \vec d + \vec h f_s.
\end{align}
In the superconductor, the solution for the anomalous Green function is given by \begin{equation}\label{eq:BCS}
    f_{BCS} = \frac{\Delta \Theta(E^2 - \Delta^2) \sign E}{\sqrt{E^2 - \Delta^2}} 
    -\i \frac{\Delta \Theta(\Delta^2 - E^2) }{\sqrt{\Delta^2 - E^2}},
\end{equation}
where $\Theta$ is the Heaviside step function and $\text{sgn}$ is the sign function. 
It is observed from these linearized equations that if $f_s = -\tilde{f}_s$, which holds for a BCS superconductor in a real gauge as seen from Eq. (\ref{eq:BCS}), then $\vec d = \tilde{\vec d}$ must hold. This relation continues to hold even in the presence of Rashba and Dresselhaus spin-orbit coupling, in which case \cite{jacobsen_prb_15} an additional term $2iDa^2\Omega(\chi){\vec d}$ appears on the lower line of Eq.~(\ref{eq:linear}). In this term, $a$ denotes the strength of the SOC, and $\Omega$ is a matrix with real entries where $\chi$ depends on the relative magnitude of Rashba and Dresselhaus coupling.
We proceed by using the Kuprianov-Lukichev boundary conditions \cite{kupriyanov_jetp_1988}, which for the upper right component of the anomalous Green function reads \cite{bergeret_rmp_05}
\begin{equation}
    \label{eq:bdc}
    \partial_x \ubar f_2  = - 2 \Omega_2 \ubar f_1
\end{equation} 
when assuming the magnitudes of the components of $\ubar f_2$ to be much smaller than the magnitude of the components of $\ubar f_1$. This approximation holds in the weak proximity regime when the non-superconducting material is in contact with a BCS superconductor at $T \ll T_c$. 
Here, 1 denotes the material to the left of the interface, and 2 the material to the right of the interface, and we introduced the boundary and material specific constant $\Omega_2 = (2 \zeta_2 L_2)^{-1}$, where $L_2$ is the length of material $2$, and $\zeta_2$ is the ratio between the bulk resistance and the interface resistance. Other types of boundary conditions can be used for highly transparent interfaces \cite{nazarov_sm_99}, magnetic interfaces \cite{cottet_prb_09, eschrig_njp_15}, or spin-orbit coupled interfaces \cite{amundsen_prb_19, linder_prb_22}.

For concreteness, consider a magnetization in the \textit{z}-direction inside the ferromagnet.
The solution for the anomalous Green function in the ferromagnet then becomes 
\begin{align}
    \begin{split}
        d_z(x) &= C_1 \e{k_- x} + C_2 \e{-k_- x} + C_3 \e{k_+ x} + C_4 \e{-k_+ x}
        \\
        &\approx C_2 \e{-k_- x} + C_4 \e{-k_+ x}
    \end{split}
    \\
    \begin{split}
    f_s(x) &= -C_1 \e{k_- x} + -C_2 \e{-k_- x} + C_3 \e{k_+ x} + C_4 \e{-k_+ x}
    \\
    &\approx -C_2 \e{-k_- x} + C_4 \e{-k_+ x},
    \end{split}
\end{align}
where we defined
\begin{align}
k = \sqrt{-\frac{2 \i E}{D}} &&
k_\pm = \sqrt{-\frac{2 \i (E \pm h)}{D}}.
\end{align}
Above, $h$ is the magnitude of the spin-splitting of the conduction bands in the ferromagnet, $\{B_j,C_1, C_2, C_3, C_4\}$ are constants, and we have assumed semi-infinite ferromagnets, such that the exponentially increasing parts of the solutions can safely be ignored. 
The solutions for magnetizations in the \textit{x}- or \textit{y}-direction are similar upon exchanging $d_z$ with $d_{x,y}$ functions. 
Applying Eq.~\eqref{eq:bdc}, we find 
\begin{align}
    C_2 =- \frac{\Omega_2 f_{BCS}}{k_-} && 
    C_4 =  \frac{\Omega_2 f_{BCS}}{k_+},
\end{align}
from which one verifies that $\vec d$ is indeed symmetric under the tilde conjugation operation, while $f_s$ obtains a sign change under the same operation. 
These symmetries follow from the fact that $\tilde k = k$ and $\tilde k_\pm = k_\mp$.

Importantly, we have also verified the property $\tilde{d}(E) = d(E)$ for the odd-frequency anomalous Green function using a full numerical solution without approximations, using the Ricatti parametrization, in S/F multilayers using a real superconducting order parameter $\Delta=|\Delta|$ in S. In the case of a complex $\Delta=|\Delta|\e{\i\phi}$, the induced odd-frequency correlator instead satisfies
\begin{equation}
    \tilde d(E) = d(E)\e{-2\i\phi},
\end{equation}
which may alternatively be written $\tilde{d}(E,\phi) = d(E,-\phi).$

\subsection{Required criteria for the Green function matrix}
Motivated by the solution in the last section, we will now establish necessary criteria that the
retarded quasiclassical Green function should satisfy in the odd-frequency case, including conserved spectral weight. Our goal will be
to find a model for a bulk odd-frequency superconductor that satisfy these properties, as well
as reproducing results that are expected for such superconductors. We assume that we are in the diffusive regime of transport, and only consider $s$-wave superconductivity, in which case triplets will be odd in
frequency and singlets will be even in frequency. \\

When restricting ourselves to a real order parameter, we showed in the previous subsection that for an S/F system with a real order parameter in S, the singlet and triplet anomalous Green functions satisfy
\begin{align}
    \tilde f_{s} &=  - f_{s}\\
    \tilde{\vec d} &=  \vec d.
\end{align}
Now, the spectral weight function $\mathcal A(\vec p, E)$ is defined from the Green function $G$ as 
\begin{equation}
    \mathcal A_\sigma(\vec p, E) = - \frac{1}{\pi} \Imag \{G^R_{\sigma \sigma}\},
\end{equation}
which when we apply the quasiclassical approximation, $G = -\i \pi \delta(\xi_p) g(\vec p_F, E)$, becomes
\begin{equation}
    \mathcal A_\sigma(\vec p, E) = \Real \{g_{\sigma \sigma} \delta(\xi_p)\},
\end{equation}
reflecting the fact that in the quasicassical approximation, the Green function describes a low-energy theory for electronic states with momentum equal to the Fermi momentum $\vecp_F$. 
This means that the standard sum rule of the spectral weight, which reads
\begin{equation}
    \label{eq:sum_rule_normal}
    \int_{-\infty}^\infty dE \mathcal A_\sigma(\vec p, E) = 1,
\end{equation}
must be modified accordingly. This sum rule states that the total number of states is conserved in the system, which holds upon transitioning into the superconducting state even though the Green function and spectral weight changes.

Comparing a general solution to the bulk normal metal solution, $\underline g = \underline 1$, we may express this criterion using the quasiclassical Green function as
\begin{equation}
\label{eq:sum_rule_criterion}
    \int^{\omega_c}_{-\omega_c} dE (N_\sigma - N_0) = 0.
\end{equation}
Formally, this expression can be obtained by integrating Eq.~\eqref{eq:sum_rule_normal} over momenta to remove the delta functions, and introducing the density of states $N_\sigma = \text{Re}\{g_{\sigma \sigma} \}$ as well as the density of states at the Fermi surface, $N_0$. 
Above, $\omega_c$ is the high-energy cutoff in quasiclassical theory, which in practice is chosen large enough to ensure that the normal-state solution for the Green function is recovered at energies close to $\omega_c$.

When deriving the Usadel equation, information about the normalization of the Green function is lost when subtracting the left-handed and right-handed equations of motion.
We will here use the standard choice of normalization $\check{g}^2 = \check{1}$, which leads to for instance 
\begin{equation}
\label{eq:normalization}
    \underline g^2 - \underline{f} \underline{\tilde f} = \underline 1.
\end{equation}
This will serve as the definition for the normal component of the quasiclassical Green function when we make \textit{ans{\"a}tze} for the anomalous component. Finally, we expect the superconducting correlations to vanish for large energies, analogous to the BCS case. 
This corresponds to a vanishing anomalous Green function for large energies. 

Summarizing, the four criteria we demand for the retarded Green function describing odd-frequency superconductivity are as follows:
\begin{itemize}
\item Energy-symmetry: 
${\vec d}(E,\phi) = \tilde{\vec d}(E,-\phi)$
\item Conserved spectral weight: $\int^{\omega_c}_{-\omega_c} dE (N_\sigma - N_0) = 0$.
\item Normalization: $\hat{g}^2 = \hat{1}$.
\item Vanishing correlations for large energies: $\lim_{|E|\to\infty} {\vec d}(E) = 0$.
\end{itemize}

\subsection{Meissner response}
Before proceeding to discuss the particular Meissner response exhibited by various anomalous Green functions that satisfy the four criteria listed above, it is useful to briefly sketch how odd-frequency pairing can fundamentally change the (orbital) magnetic response of a superconductor.

In the London gauge, the Maxwell equation for a time-independent system reads
\begin{equation}
        \vec \nabla^2 \vec A = - \vec j,
\end{equation}
where $\vec A$ is the magnetic vector potential and $\vec j$ is the electric current density.
Inserting the London equation, $\vec j = - \vec A / \lambda_L$, where $\lambda_L$ is the London penetration depth of the system, one finds from the solution of the Maxwell equation an exponentially decaying magnetic field, consistent with the Meissner effect in conventional superconductors. 

The current density response to a magnetic field in a bulk superconductor can be written as
\begin{equation}
    \vec j = 
    -\frac{\i \vec A N_0 e^2 D}{16} \int_{-\infty}^\infty dE\,  \Tr{ \hat \tau_3 \left(\check g  \bigr[ \hat \tau_3, \check g \bigl]\right)^K
    },
\end{equation}
where the “K” superscript denotes the upper right component in Keldysh space.

This can be simplified considerably by assuming an equilibrium situation,
\begin{equation}
    \vec j =  \vec A \frac{N_0 e^2 D}{2} I(x),
\end{equation}
where we introduced the temperature-dependent integral
\begin{equation}
    \label{eq:I_integral}
    I(x) =  \int_{-\infty}^\infty dE\, \tanh\! 
    \left(\frac{E}{2T}\right) 
    \Tr {\Imag \bigl\lbrace \underline f \underline{\tilde f}  \bigr\rbrace },
\end{equation}
which will determine the sign of the current.

In the case of an unconventional Meissner response where $\vec j =  \vec A / \lambda_L$, the solution for the magnetic field becomes oscillating \cite{yokoyama_prl_11, dibernardo_prx_15}.
We now consider a one-dimensional superconducting system in the \textit{x}-direction, subject to an external magnetic field, such that the values at the boundaries become $\vec B = B_0 \vec e_z$.
Gauge freedom then allows us to choose $\vec A = A_y(x) \vec e_y$.
The system length is $L$, producing the boundary conditions
\begin{align}
    \frac{dA_y}{dx}(x = 0) = B_0 && \frac{dA_y}{dx}(x = L) = B_0.
\end{align}

\begin{figure}[htb]
    \centering
    \includegraphics[width=6.cm]{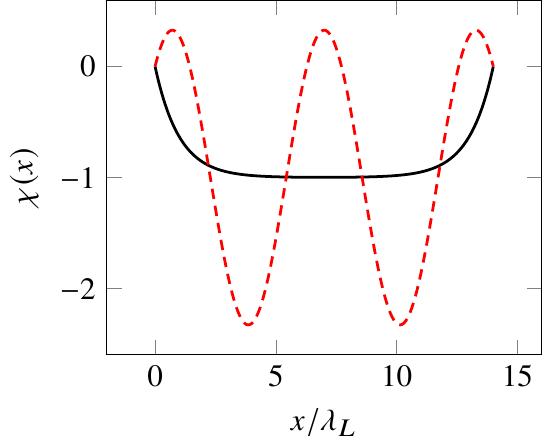}
    \caption{The magnetic susceptibility $\chi$ of a material with length $L = 14 \lambda_L$ with a conventional (solid) and unconventional (dashed) Meissner response. 
    The values of $\chi < -1$ are unphysical.}
    \label{fig:magnets}
\end{figure}

It can then be shown that the conventional Meissner response produces a field 
\begin{equation}
    \label{eq:magnetic_field_conv}
    B_z(x) = B_0 \frac{\e{ \lambda_L^{-1} L} - 1}{  \e{ \lambda_L^{-1} L} - \e{- \lambda_L^{-1} L}} \e{- \lambda_L^{-1} x} +
    B_0 \frac{\e{- \lambda_L^{-1} L} - 1}{  \e{ \lambda_L^{-1} L} - \e{- \lambda_L^{-1} L}} \e{\lambda_L^{-1} x},
\end{equation}
while the unconventional response produces 
\begin{equation}
    B_z(x) = B_0 \cos(\lambda_L^{-1} x) - B_0  \frac{\cos( \lambda_L^{-1} L) - 1}{\sin( \lambda_L^{-1} L)} \sin(\lambda_L^{-1} x).
\end{equation}
Defining the magnetic susceptibility
\begin{equation}
    \chi(x) = \frac{B_z(x)}{B_0} - 1,
\end{equation}
we plot in Fig.~\ref{fig:magnets} the two responses. 
We clearly observe that the unconventional Meissner response produces an oscillating susceptibility, corresponding to both a local paramagnetic and diamagnetic response.

The energy density stored in the magnetic field is $\vec B^2/(1 + \chi)$, meaning that for values $\chi< -1$, we find that the energy density becomes negative, indicating an instability of the system \cite{yokoyama_prl_11}. Another way to see that the paramagnetic Meissner effect is problematic for a bulk superconductor is by examining the Landau expansion for the Gibbs free energy:
\begin{align}
F &= \int \Big[ \alpha(T) |\psi|^2 + \frac{1}{2}\beta(T)|\psi|^4 + \gamma(T) |(\nabla + 2\i e/\hbar c) \psi|^2 \notag\\
&+ B^2/8\pi\Big]d^3x.
\end{align}
This free energy must be minimized with respect to variations of $\psi$ and $\vecA$. Minimization with respect to $\vecA$, $\delta F/\delta \vecA=0$, gives
\begin{align}
\nabla\times \vecB = (4\pi/c)\boldsymbol{J},
\end{align}
where $\boldsymbol{J}$ is proportional to the supercurrent:
\begin{align}
\boldsymbol{J} = -(4e\gamma/\hbar)|\psi|^2 (\nabla\phi + 2e\vecA/\hbar c).
\end{align}
For $\gamma > 0$, we have a diamagnetic Meissner effect and the London penetration depth is
\begin{align}
\lambda_L^{-2} = 32\pi(e^2/\hbar^2c^2) \gamma|\psi|^2.
\end{align}
The superfluid density is then $n_s \propto \gamma|\psi|^2$. The above equations demonstrate why a paramagnetic Meissner effect in a homogeneous superconductor becomes thermodynamically unstable: to reverse the supercurrent response due to $\vec A$, we must set $\gamma < 0$. 
This makes the superfluid density negative. 
The problem is now that the free energy term $\propto\gamma$ can be made increasingly negative by making $\nabla\psi$ larger and larger in magnitude. 
There is no lower bound as $\psi$ becomes ever more inhomogeneous, which is unphysical. 

A bulk intrinsic odd-frequency superconductor has yet to be discovered experimentally. In such a system, one would expect the Meissner response to be diamagnetic in order for the superconducting phase to be thermodynamically stable. Note that this is not required for the odd-frequency superconductivity induced in heterostructures, such as S/F bilayers, which can exhibit paramagnetic behavior. 
The coexistence of odd-frequency correlations displaying a positive and negative superfluid density, respectively, has shown to be unphysical \cite{fominov_prb_15}. Several works have shown that a diamagnetic odd-frequency superconducting state is thermodynamically stable \cite{solenov_prb_09, kusunose_jpsj_11}, while Ref. \cite{fominov_prb_15} raised concerns about the existence of a mean-field Hamiltonian describing such a state. For a detailed discussion, see Ref. \cite{linder_rmp_19}.
The fact that the Meissner response can be either diamagnetic, paramagnetic, or even absent depending on which particular \textit{ansatz} one chooses for the odd-frequency anomalous Green function $f$ emphasizes the care that must be taken when considering effective models for odd-frequency superconductivity.

\section{Results and Discussion}

There are several ways that the condition $\vec d = \tilde{\vec d}$ can be satisfied, since the $\tilde{\ldots}$-operation involves both inversion of energy $E \to -E$ and complex conjugation. However, choosing an anomalous Green function which satisfies this symmetry can still lead to physically problematic behavior, such as non-conserved spectral weight, as in Ref. \cite{tanaka_prl_07}.
Below, we give three examples which fulfill the four criteria listed previously, yet all have different magnetic responses. For simplicity, we consider only the $S_z=0$ triplet component $d_z$ in what follows. 

\subsection{Model I: diamagnetic Meissner response}
We consider a model
\begin{align}
    d_z= A(E)  f_{\text{BCS}},
\end{align}
where $A$ is a real odd function of energy that must be chosen such that Eq.~\eqref{eq:sum_rule_criterion} holds.
Using normalization, the normal Green function becomes 
\begin{equation}\label{eq:gmodelI}
    \underline g 
    = \sqrt{\frac{E^2 - \Delta^2(1 -  [A(E)]^2)}{E^2 - |\Delta|^2}} \underline 1.
\end{equation}

In order to calculate the Meissner response in this model, we consider the integral from Eq.~\eqref{eq:I_integral}, starting with the trace,
\begin{equation}
    \Tr{\Imag \lbrace \underline f \tilde{ \underline f} \rbrace} = \frac{- 2 [A(E)]^2 |\Delta|^2 \sign{E} \delta'}{\delta'^2 + (E^2 - |\Delta|^2)^2},
\end{equation}
where we have used that $E\rightarrow E + \i \delta$ in the retarded Green function, where $\delta \in \mathbb{R}$ is a positive infinitesimal, and introduced $\delta' = 2|E| \delta$.
Evaluating this Lorentzian as $\delta' \rightarrow 0$ and performing the integral, we find that the current becomes
\begin{equation}
    \vec j = - [A(\Delta)]^2 \frac{\vec A}{\lambda_L},
\end{equation}
where $\lambda_L$ is the \text{BCS} proportionality constant.
This is a conventional diamagnetic Meissner response with a renormalized London penetration depth given in terms of the square of the antisymmetric function.
As a simple example, one could consider a function $A(E) = \alpha \sign{E}$, which for the choice $\alpha = 1$ would produce the exact same Meissner response as in the BCS case.
Moreover, it can be seen that the density of states is equivalent to the BCS density of states.

Initially, this result may seem at odds with the general result that systems with odd-frequency pairing should have a density of states larger or equal to one at $E = 0$, which is derived in Appendix A. This apparent paradox is understood by noting that 
the sign function is defined so that $\sign{E=0}=0$, ensuring $[A(E)]^2 = 0$ for $E= 0$.
This causes the density of states to be nonzero in an infinitely narrow peak centered at $E = 0$, as seen from Eq. (\ref{eq:gmodelI}) when setting $A(E=0)=0$.
Since this peak is infinitely narrow, while the height of the peak is finite, the contribution to the spectral sum rule vanishes, and we can ignore this peak in the following. The discussion above holds also for other choices of $A(E)$ that produce a gap at the Fermi energy.
This function can also be shown to preserve the sum rule of the spectral weight for $\alpha \leq 1$.
For the special case of $\alpha = 1$, the density of states have the exact same form as for a BCS superconductor.

\begin{figure}[htb]
    \centering
    \includegraphics[width=6.cm]{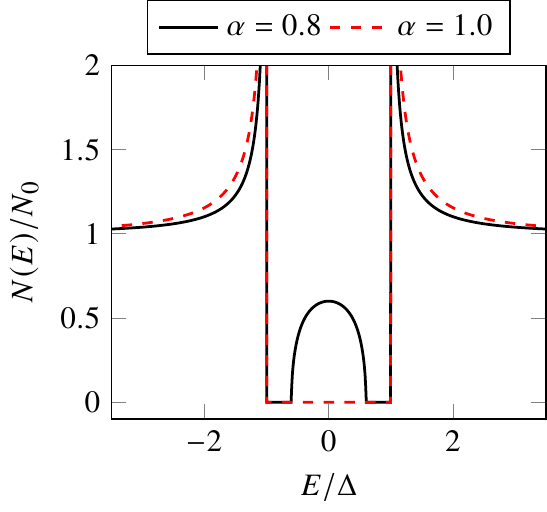}
    \caption{The density of states for Model I for different values of $\alpha$.}
    \label{fig:dos_model1}
\end{figure}

\subsection{Model II: mixed Meissner response}
Next, we study the model
\begin{align}\label{eq:modelII}
    d_z = \i S(E) f_\text{BCS},
\end{align}
 where $S$ is a real even function of energy that must be chosen such that Eq.~\eqref{eq:sum_rule_criterion} holds.
The normal Green function of the system is given through the normalization condition as
\begin{equation}
    \underline g(E) = \sqrt{\frac{E^2 -\left(1 + [S(E)]^2\right) \Delta^2}{E^2 - \Delta^2}} \underline 1,
\end{equation}
choosing the positive root in order for the system to reduce to the normal metal case when the order parameter vanishes.

Some general remarks can be made about $\underline g$ and the corresponding density of states, even without specifying the function $S$.
The density of states is peaked at zero energy, and is larger than unity for $E < \Delta$.
Additionally, $\underline g$ is purely real for energies in the domains $|E| < |\Delta|$ and $|E| > \sqrt{1 +  S(E)} |\Delta|$, and purely imaginary (thus providing a gapped density of states) in between.
This means that if one wants a model without a gap in the density of states, $S$ must be chosen in such a way that it satisfies $ S(\Delta) = 0$.
For this model to satisfy all of our criteria discussed above, we only need to choose a function $S(E)$ that causes the number of states to be conserved for all choices of $\Delta$.
The easiest such choice is $S(E) = s \in \mathbb{R}$.
Using elliptic integrals, it can be shown that this choice makes the Green function satisfy the sum rule of the spectral weight.
This continues to holds for any $\Delta$, because $f_{BCS}$, and thus $d_z$ in this case, depends on $\Delta$ only through the ratio $E/\Delta$.

\begin{figure}[htb]
    \centering
    \includegraphics[width=6.5cm]{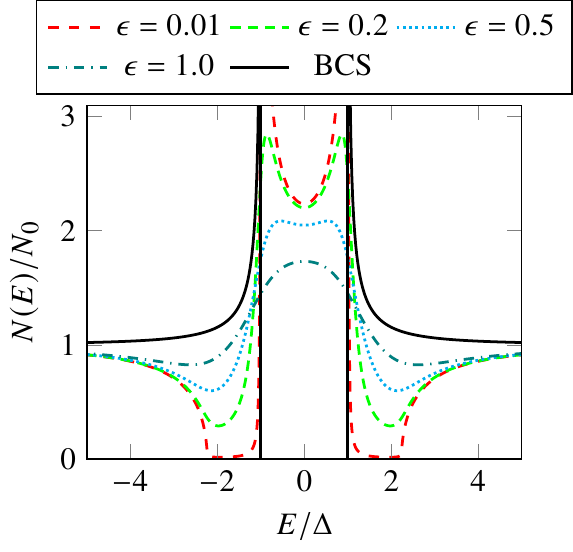}
    \caption{The density of states for Model II with $S(E) = 2$ when adding a Dynes \cite{dynes_prl_78} inelastic scattering parameter $\epsilon \equiv \delta/\Delta$. The BCS density of states with a Dynes parameter $\epsilon = 0$ is also included for comparison.
    }
    \label{fig:dos_model2}
\end{figure}

In Fig~\ref{fig:dos_model2}, the density of states of this model is plotted along with the BCS density of states for comparison. We have included the effect of inelastic scattering via a Dynes parameter~\cite{dynes_prl_78} $E \to E +\i\delta$ as in the previous section, since it is always present to some extent in actual experiments. We have also verified numerically that adding inelastic scattering in this way does not cause a violation of Eq. (\ref{eq:sum_rule_criterion}). As seen in Fig.~\ref{fig:dos_model2}, a very large amount of inelastic scattering $\delta \simeq \Delta$ is required to make the density of states have a clear peak at the Fermi level for this particular model. 
However, we remark that the $\delta$ appearing in $d_z = \i S(E) f_\text{BCS}$ in this way does not need to correspond to actual inelastic scattering: alternatively, $\delta$ may be considered a fitting parameter to give the anomalous Green function the desired density of states behavior.

By a similar calculation as in the last subsection, we find that 
\begin{equation}
    \vec j = [S(\Delta)]^2 \frac{\vec A}{\lambda_L},
\end{equation}
which is an unconventional Meissner response, producing an oscillating susceptibility.
For the special case of $S(E) = 1$, this is the exact response discussed in the last section.

\subsection{Model III: no Meissner response}
For completeness, we finally include a model for the odd-frequency anomalous Green function which does not give rise to any Meissner response. Although unphysical, we include this for the purpose of illustrating that writing down an \textit{ansatz} for the anomalous Green function which satisfies the four criteria outlined previously, including conservation of spectral weight, is not sufficient to produce a Meissner response, be it diamagnetic or paramagnetic. 
The \textit{ansatz} reads
\begin{align}
d_z = c\theta(a^2-E^2) + \i d \text{sgn}(E) \theta(E^2-a^2)\theta(b^2-E^2),
\end{align}
where $a,b,c, d$ are positive real numbers.
The normal Green function becomes
\begin{equation}
    \ubar g = g \ubar 1,
\end{equation}
with
\begin{equation}
        g 
        =   c_2 \Theta(a^2- E^2)  + d_2 \Theta(E^2 - a^2 )\Theta(b^2 - E^2) + \Theta(E^2 - b^2),
\end{equation}
where we introduced the auxiliary parameters $c_2 \equiv \sqrt{1 + c^2}$ and $d_2 \equiv \sqrt{1 - d^2}$.

\begin{figure}[htb]
    \centering
    \includegraphics[width=6.cm]{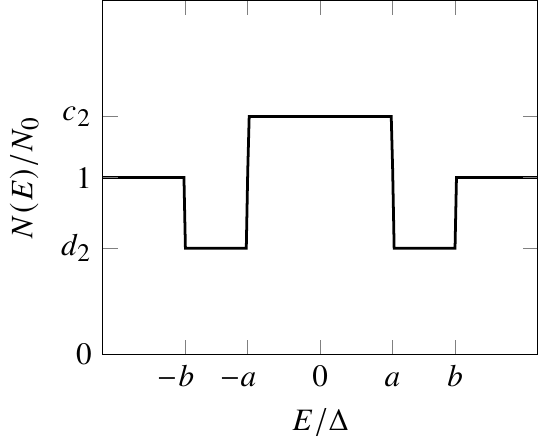}
    \caption{The density of states for Model III. }
    \label{fig:dos_model3}
\end{figure}
This leads to a peaked density of states, which is plotted in Fig. \ref{fig:dos_model3}. 
By fixing one of the parameters, it can be shown to satisfy the sum rule. It is, however, readily shown that 
\begin{equation}
    \Imag \{\underline f \tilde{\underline f}\} = 0,
\end{equation}
which causes the Meissner response to vanish for this choice of effective odd-frequency anomalous Green function. 

\subsection{Model IV: numerical solution in full proximity effect regime}
The models considered above (I-III) give simple expressions for the density of states and Meissner response. 
\begin{figure}[htb]
    \centering
    \begin{subfigure}[t]{0.49 \textwidth}
        \includegraphics[width = 6.cm]{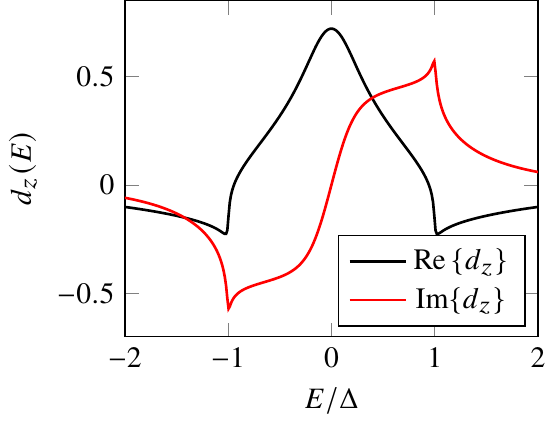}
        \caption{The real and imaginary part of the anomalous Green function for Model IV at the vacuum edge.}
        \label{fig:realim_end}
    \end{subfigure}
    \hfill
    \begin{subfigure}[t]{0.48 \textwidth}
        \includegraphics[width = 6.cm]{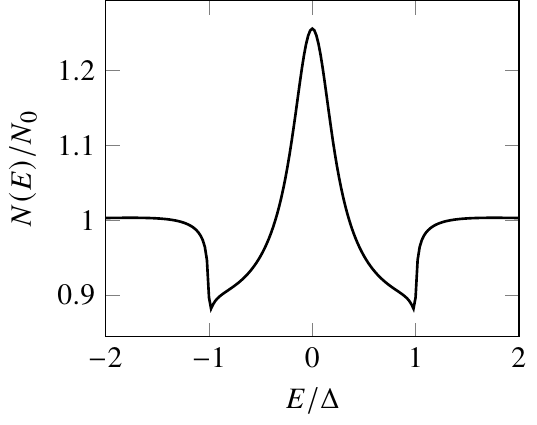}
        \caption{The normalized density of states for Model IV at the vacuum edge.
        }
     \label{fig:dos_end}
    \end{subfigure}
    \caption{The proximity effect for Model IV at the vacuum edge.
    Parameters used are $h = 50 \Delta$, $\Omega = 6.7 \xi^{-1}$, $L = 0.75 \xi$, and $\epsilon = 0.01$.}
    \label{fig:end_fullsys}
\end{figure}
It is instructive to compare these with the properties of a system where odd-frequency superconductivity arises naturally through the proximity effect. Here, we consider an S/F/F/N system, modelled with a BCS equilibrium superconductor in contact with a single ferromagnet with a varying magnetization through an interface with a boundary specific constant $\Omega$.
The origin ($x = 0$) is at the S/F interface.
A natural length scale in dirty superconducting systems is the superconducting coherence length $\xi = \sqrt{D / \Delta}$, which we use to make the Usadel equation dimensionless. 
Lengths are given in units of $\xi$, and energies in units of $\Delta$.
The solutions for the quasiclassical Green functions were found using the Riccati parametrization \cite{maki_prb_95, konstandin_prb_05}.
The magnetization close to the interface is pointing in the \textit{z}-direction, and, over a short distance, rotates with a uniform rate into pointing in the \textit{x}-direction. Throughout the ferromagnet, the spin-splitting $h$ due to the magnetization has a constant magnitude. At the F/N interface, the spin-splitting is set to decay exponentially over a short distance.
Tuning the parameters allows us to make a system where the triplet amplitude dominates in the normal metal part of the material. In this way, the normal metal becomes an effective odd-frequency superconductor.
We consider the vacuum end of the normal metal, at $x = L$, and plot the real and imaginary part of the triplet component $d_z$ in Fig. \ref{fig:realim_end}, as well as the density of states in Fig. \ref{fig:dos_end}. As seen, the energy-dependence of the real and imaginary part is not easily described by a simple trial function.

\begin{figure}[h]
    \centering
    \begin{subfigure}[t]{0.49 \textwidth}
        \includegraphics[width = 6.cm]{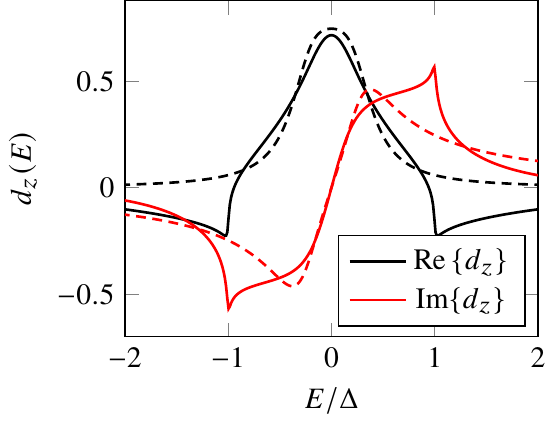}
    \end{subfigure}
    \hfill
    \begin{subfigure}[t]{0.48 \textwidth}
        \includegraphics[width = 6.cm]{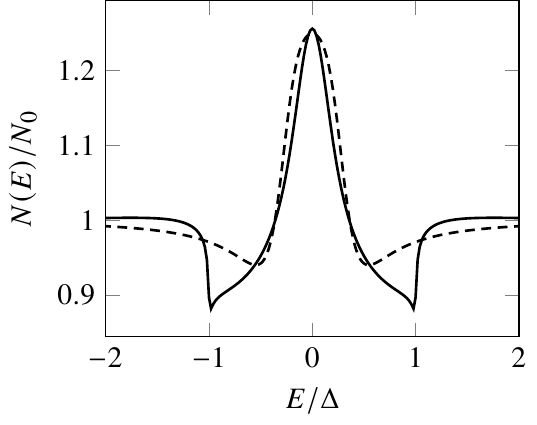}
    \end{subfigure}
    \caption{A qualitative fit of Model II (dashed lines) with a constant function $S(E)$ to the results from the numerically exact solution of the S/F/F/N junction (Model IV) (solid lines), with the same parameters as in Fig \ref{fig:end_fullsys}.
    The parameters $\{S(E), \epsilon_f,\Delta_f\}$ used in Model II, to be distinguished from the parameters $\{\epsilon,\Delta\}$ used in the numerical solution of Model IV, to make the fit are $S(E) = 0.88 $, $\epsilon_f =0.19 $, and $\Delta_f = 0.25\Delta$.}
    \label{fig:fit}
\end{figure}
Moreover, we calculated the Meissner response in the S/F/F/N proximity system and found that close to the superconductor, where singlet components dominate, the Meissner response is conventional. In the N part of the system, where odd-frequency pairing dominates, the Meissner response is unconventional.
The peaked density of states at the Fermi energy as well as the sign of the Meissner response suggests that the function could be fitted with Model II above. However, the non-trivial form of the solution shows that a complicated function $S(E)$ must be chosen in order to get a very good quantitative fit. Nevertheless, a qualitative and quantitatively comparable fit can be obtained by simply choosing $S(E)$ to be a constant as well as letting the Dynes parameter take the value $\epsilon_f$ and the order parameter take the value $\Delta_f$ as curve fit parameters, which in general can take different values than the corresponding quantities $\epsilon, \Delta$ in the actual proximity system.
Using Eq. (\ref{eq:modelII}), a fit of the anomalous Green function and the density of states for the S/F/F/N proximity system are shown in Fig.~\ref{fig:fit}. 
Numerically, it can be verified that the density of states satisfy Eq.~\eqref{eq:sum_rule_criterion}. It is interesting to note the compact form of the effective odd-frequency anomalous Green function $d_z$ in Eq. (\ref{eq:modelII}) used to fit the actual numerical solution of the S/F/F/N system: it is simply proportional to the BCS anomalous Green function, which makes it useful to work with practically.

\section{Summary}

In this work, we establish a set of necessary criteria that a physically sound quasiclassical Green functions describing odd-frequency superconductivity in the dirty limit must satisfy. We demonstrate that it is possible to write down effective models for the Green function which satisfy all the abovementioned criteria, but which nevertheless give rise to different behavior when it comes to the density of states and the magnetic response of the superconductor.
This highlights the importance of carefully considering the properties of effective models describing odd-frequency superconductivity and may provide a guide for future works considering emergent phenomena associated with odd-frequency superconductors.

\begin{acknowledgments}
We thank S. Aunsmo, J. A. Ouassou, and L. G. Johnsen for useful discussions and A. Golubov for correspondence. We acknowledge funding via the Research Council of Norway Grant numbers 323766, as well as through its Centres of Excellence funding scheme, project number 262633. J. L. also acknowledges support from the Sigma2 project no. NN9577K.
\end{acknowledgments}

\appendix

\section{Proof of increased (decreased) density of states at $E=0$ for triplets (singlets)}

In this appendix, we will derive a relation describing how the symmetries of the quasiclassical Green function determine the effect of the density of states at the Fermi energy.
Specifically, we will show that for triplet odd-frequency pairing, the density of states must be larger or equal to one, while for singlet even-frequency, it must be lowered. 
Consider a system with both singlet and $d_z$ triplet pairing, described by 
\begin{equation}
    \ubar{f} = \begin{pmatrix}
        0 & d_z + f_s\\
        d_z - f_s & 0 
    \end{pmatrix},
\end{equation} 
and consider $E = 0$, meaning that $\tilde f(0) = [f(0)]^*$. 
The proofs for the other triplet components are completely analogous. 
From the normalization condition we get, temporarily suppressing the $E = 0$ argument, and introducing the quantity $c \equiv fs^* d_z -f_s d_z^*$,
\begin{equation}
    \underline g^2= 
    \begin{pmatrix}
    1 + |d_z|^2 - |f_s|^2 + 2\i \Imag \{c\} & 0\\
    0 &   1 +  |d_z|^2 - |f_s|^2 - 2\i \Imag\{c\} 
    \end{pmatrix}.
\end{equation}
Taking the positive square root of this equation yields an expression for the normal part of the quasiclassical Green function.
In order to get the spin averaged density of states, we average over the spin-up and spin-down pairing density, and take the real part of the quasiclassical Green function, 
\begin{equation}
    \label{dos_0_general}
    N(E = 0) =\frac{N_0}{2} \text{Re} \{g_{\uparrow \uparrow} + g_{\downarrow \downarrow}\}.
\end{equation}
In the case of a weak proximity effect, we expand the square root to first order, and get
\begin{equation}
    N(E = 0) = N_0 \left(1 + \frac{|d_z|^2 - |f_s|^2}{2} \right).
\end{equation}
This shows that in the weak regime, singlets will always lower the zero-energy density of states, while triplets will increase it. This property was also noted in Ref. \cite{kawabata_jpsj_13} in the weak proximity effect regime.

In the case of a strong proximity effect, we cannot expand the square root, and the expression will not reduce in the same way as in the weak proximity regime. 
If we, however, consider systems with only singlets or only triplets, we find that 

\begin{equation}
    \label{eq:dos_result}
    N(E = 0) =
    \begin{cases}
        N_0 \sqrt{1 - |f_s|^2}, & \text{ even-frequency,}\\
        N_0 \sqrt{1 + |d_z|^2}, & \text{ odd-frequency}.
    \end{cases}
\end{equation}
We thus see that also in the full proximity effect regime, the triplets increase the density of states at zero energy, while the singlets decrease it.

\end{document}